\documentclass[showpacs,reprint,nofootinbib,longbibliography]{revtex4-1}
\usepackage[utf8x]{inputenc}
\usepackage{amsmath}
\usepackage{amsfonts}
\usepackage{amssymb}
\usepackage{graphicx}
\usepackage{bbm}
\usepackage{xspace}
\usepackage{verbatim}
\usepackage[normalem]{ulem}
\usepackage{enumitem}
\usepackage{hyperref}

\setlist{topsep=2pt,itemsep=0pt}
\usepackage{cancel}

\newcommand{\SU}{\text{SU}}

\newcommand{\dd}{D}

\newcommand{\eo}{e}
\renewcommand{\oe}{o}

\newcommand{\one}{\mathbbm{1}}

\newcommand{\Gconj}[1]{\overline{#1}}

\DeclareMathOperator\arsinh{arsinh}
\DeclareMathOperator\tr{tr}

\newcommand{\ep}{e^{\mu}}
\newcommand{\epp}{e^{3\mu}}
\newcommand{\emu}{e^{-\mu}}
\newcommand{\emum}{e^{-3\mu}}

\newcommand{\hi}{\hat{i}}

\usepackage{color}

\newcommand\reduline{\bgroup\markoverwith{\textcolor{red}{\rule[0.5ex]{2pt}{0.4pt}}}\ULon}

\newcommand{\JBwave}{\bgroup \markoverwith{\color{red}\lower4pt\hbox{\sixly\char58}}\ULon}

\newcommand\fbuline{\bgroup\markoverwith{\textcolor{blue}{\rule[0.5ex]{2pt}{0.4pt}}}\ULon}

\newcommand{\ub}{\bar u}
\newcommand{\vb}{\bar v}
\newcommand{\psiy}{\varphi}
\newcommand{\Nt}{N_t}
\newcommand{\Nmc}{N_\text{MC}}

\begin{document}

\title{Positivity of center subsets for QCD}

\author{Jacques Bloch}
\email{jacques.bloch@ur.de}
\author{Falk Bruckmann}
\email{falk.bruckmann@ur.de}
\affiliation{Institute for Theoretical Physics, University of
  Regensburg, 93040 Regensburg, Germany}
\date{November 13, 2015}


\begin{abstract}
We further pursue an approach to the sign problem of quantum chromodynamics at nonzero chemical potential, in which configurations of the lattice path integral are gathered into subsets. In the subset construction we multiply each temporal link by center elements independently and in a first step neglect the gauge action. The positivity of the subset weights -- shown for 0+1 dimensions in an earlier study -- extends to larger lattices: for two sites in the temporal direction and arbitrary spatial extent we give a proof of the positivity by decomposing the subset weight in positive summands. From numerical evidence we conjecture that the positivity persists on larger lattices and that the gauge action can be reintroduced through mild reweighting. First results on the quark number obtained with this method in two dimensions are shown as well.
\end{abstract}

\maketitle

\section{Introduction}

The sign problem in quantum chromodynamics (QCD) -- the non-positivity of the quark determinant at nonzero chemical potential -- hampers numerical simulations of QCD thermodynamics based on importance sampling, see e.g.\ \cite{deForcrand:2010ys}. This problem, generically caused by a complex action in the partition function, also occurs in other physical systems \cite{Gattringer:2014nxa,Aarts:2015kea}. The sign problem causes large cancellations between the contributions of the configurations corresponding to the fundamental degrees of freedom of the ensemble, and suggests that other degrees of freedom may be more effective to render the path integral in such regimes.

The subset method consists of gathering configurations of the ensemble into subsets using some definite rule. The weight of such subsets is given by the sum of the individual weights. An `early application of subsets' (to summations) is the computation of finite arithmetic series by first pairing up numbers positioned symmetrically around the center of the sequence and then summing up the pair sums (this goes back to the Indian mathematician Aryabhata \cite{Aryabhat499AD} and is also attributed to the young C.F.~Gauss). In lattice QCD and related theories the aim is to find subsets with positive weights. The remaining path integral over all subsets can then be treated by statistical methods like importance sampling. 

The subset method has been first developed on a random matrix model of QCD: subsets with positive weights could be found, hence, solving the sign problem \cite{Bloch:2011jx,Bloch:2012ye}. Later, this subset construction was understood in terms of imaginary chemical potentials and canonical partition functions \cite{Bloch:2012bh}. Canonical partition functions vanish, when the corresponding configurations average out in the  integral. The subset method removes such contributions already on the level of the integrand, rendering the weight positive. 

Although gathering configurations in subsets and adding up their weights is a very general and always exact option to compute partition functions, there is generically no reason that the resulting subset weight be positive. In this context it is very helpful, if the generation of the configurations in the subset is guided by some symmetry principle or physical insight to perform cancellations explicitly using more physical degrees of freedom. For the random matrix model this symmetry is a $Z_N$ subgroup of the U(1)-symmetry of the integration measure. For lattice QCD, multiplying any of the group-valued links by another group element is possible\footnote{Multiplying gauge links by generic group elements has been utilized on top of the center subsets in \cite{Bloch:2013ara} to treat many flavors.}, i.e.\ compatible with the Haar measure. A natural choice, however, is to make use of the center symmetry $Z_3$ as a subgroup of the gauge symmetry SU(3), on the temporal lattice links.\footnote{Changing a single 
link and thereby changing the configurations' weights, as we do here, should not be confused with a gauge transformation that always changes several links and preserves all gauge invariants.} Mathematically speaking we divide out the center from these links. This leads to temporal mesonic and baryonic hoppings. As another motivation we take the fact that gauge groups without the center, like SU($N$) in the adjoint representation or $G_2$, do not cause a sign problem \cite{Kogut:2000ek,Hands:2000ei,Maas:2012wr}.  

In our previous work with T.~Wettig \cite{Bloch:2013ara} center subsets have been shown to yield positive weights in one of the simplest QCD toy models, QCD in 0+1 dimensions, where only the Dirac determinant is present (and where analytic solutions are available \cite{Bilic:1988rw,Ravagli:2007rw}). Since the configurations in this system are fully characterized by one group element, the Polyakov loop, the subset method consists of a single $Z_3$-multiplication. This again removes canonical determinants that average out in the partition function, and only multiples of the baryon chemical potential survive.

In the present work we extend the subset method to higher dimensions, using (unrooted) staggered fermions and in a first step neglecting the gauge action. We demonstrate that the method yields positive weights, sums of Dirac determinants, when each temporal link is multiplied by an independent center element. We prove the positivity of these center subset weights analytically for lattices with two sites in the temporal direction and arbitrary spatial extent and present numerical evidence that positivity persists for larger lattices. We show first numerical evidence that the gauge action can be reintroduced and treated through reweighting. As a first measurement using this method we present the quark number for the massless case on two-dimensional lattices of various sizes.

As it stands the subset method requires a cost growing exponentially with the volume. In the positivity proof, however, positive subterms can be identified. This could be the starting point of a refined subset method. In a companion paper \cite{Bloch:2015zzz} we give a diagrammatic representation of the subset weights shedding more light on the terms causing the sign problem and how the latter is solved by the subset method.

This paper is organized as follows: in the next section we introduce the general idea of subsets and show that they improve the reweighting factor in general. Sec.~\ref{sec:subsets_qcd} contains the definition of subsets for lattice QCD, its basic properties and building blocks. In Sec.~\ref{sec:positivity} we prove the positivity of the subset weights for lattices with two sites in the temporal direction by expressing the fermion action in terms of Grassmannians, and after that for the massless $2\times 2$ lattice using Dirac matrix language. We also comment on the possibility to apply subsets on spatial links. In Sec.~\ref{sec:num_results} we present our numerical results, which support the conjecture that this positivity also holds for larger lattices, and show data for the quark number density in two dimensions. We also give first results for the subset reweighting factor in the presence of a gauge action. Finally, we summarize and give several technical results in appendix.

\section{Idea of subsets and improvement of the reweighting factor}
\label{sec:subsets}

We consider a general integral $Z=\int\!d\mu(x)\, f(x)$ whose integrand $f(x)$ is real, but not necessarily positive. The idea is to collect several configurations $x$ and add up their weights $f(x)$ to a new weight. To formalize this we assume that the integration measure is invariant under the action of a discrete group $G$, $\int\!d\mu(x)\, f(gx)=\int\!d\mu(x)\, f(x)$ for all $g\in G$. 
For the integration over gauge groups in lattice QCD, $d\mu(U)$ is the Haar measure, which obeys this invariance. The subsets $\Omega$ generated by such a group and their weights $\sigma$ read,
\begin{align}
 \Omega_x
 &=\{gx|g\in G\}\,,\\ 
 \sigma(\Omega_x)
 &=\frac{1}{|G|}\sum_{y\in \Omega_x} f(y)
 =\frac{1}{|G|}\sum_{g\in G} f(gx)\,,
\end{align}
where $|G|$ is the cardinality of $G$ (the number of elements in $G$ and thus in $\Omega_x$ for all $x$). We divide by this number to avoid a $|G|$-fold overcounting of the configurations in the integral. 

In the best case, the new weights $\sigma(\Omega_x)$ are positive for all subsets $\Omega_x$. It means that having performed part of the integration (the summation over $y$ in $\Omega_x$) explicitly and deterministically, the remaining integral (over $\Omega_x$) may be subject to importance sampling methods. 

Even if not becoming strictly positive, the integrand always `comes closer to positivity' in the sense of an improved sign quenched reweighting factor. Besides the invariance of the measure we only need the Cauchy-Schwarz inequality to show that 
\begin{align}
 &\int \!d\mu(x)\, \big|\sigma(\Omega_x)\big| 
 = \int \!d\mu(x)\,\Big|\frac{1}{|G|}\sum_{g\in G}f(gx)\Big|  \nonumber\\
 &\leq \int \!d\mu(x)\,\frac{1}{|G|}\sum_{g\in G}|f(gx)|
 = \frac{1}{|G|}\sum_{g\in G}\int \!d\mu(x)\,|f(gx)|   \nonumber\\
 &= \frac{1}{|G|}\sum_{g\in G}\int \!d\mu(x)\,|f(x)|
 = \int \!d\mu(x)\,|f(x)|\,. 
\end{align}
For the reweighting factors $r$ and their variances $\Delta^2$ we obtain the inequalities
\begin{align}
 r_{\text{subsets}}
 \equiv\frac{Z}{\int\!dx\, \big|\sigma(\Omega_x)\big|}
 &\geq\frac{Z}{\int \!dx\,|f(x)|}\equiv r_{\text{sign quenched}},\nonumber\\
 \Delta^2_{r,\,\text{subsets}}
 &\leq \Delta^2_{r,\,\text{sign quenched}} ,
\end{align}
where the second inequality follows from the first one and the fact that $\Delta_r^2=1-r^2$ (in this case) \cite{Bloch:2012ye}. This improvement is in some sense expected, since summing/integrating over the whole ensemble yields the partition function, which shall be positive.

Note that this reduction of the sign problem comes at the expense of an increase of the computational effort by a factor of $|G|$.

\section{Subsets for lattice QCD}
\label{sec:subsets_qcd}

\subsection{Definitions}

A lattice QCD configuration is given by the tuple of temporal and spatial SU(3) links, which we denote as $\mathcal{U}=[\mathcal{U}_0,\mathcal{U}_s]$.  The partition function
\begin{align}
Z=\int\!d[\mathcal{U}_0,\mathcal{U}_s]\, \det\dd([\mathcal{U}_0,\mathcal{U}_s])
\label{ZQCD}
\end{align}
is a path integral with Haar measure $d[\mathcal{U}_0,\mathcal{U}_s]=\prod_x d\mu_H(U_0(x))\prod_{i,x} d\mu_H(U_i(x))$. Herein we work in the strong coupling limit and neglect the gauge action (in Sec.~\ref{sec:num_results} we reintroduce the gauge action through reweighting). We consider the staggered Dirac operator for one quark flavor with mass $m$ and chemical potential~$\mu$,
\begin{align}
 &\dd([\mathcal{U}_0,\mathcal{U}_s];x|y)\nonumber\\
 &=\ep U_0(x)\delta_{x+\hat{0},y}(-1)^{\delta_{x_0,\Nt}}
 -\emu U_0^\dagger(y)\delta_{x-\hat{0},y}(-1)^{\delta_{y_0,\Nt}}\nonumber\\
 &+\sum_{i=1}^{d-1}\eta_i(x)\big[U_i(x)\delta_{x+\hi,y}-U_i^\dagger(y)\delta_{x-\hi,y}\big]
 \!+\!2m \one_3 \delta_{x,y},
\label{eq:Dirac_op}
\end{align}
where we have set the lattice spacing to unity, $a=1$, and neglected a factor $2$ on the left hand side, since in the determinant of $\dd$ this only gives an irrelevant constant factor.
We work on an $\Nt\times N_1\times\ldots\times N_{d-1}$ lattice with even $N_\nu$ and antiperiodic boundary conditions in the temporal direction (represented by the minus signs on the last time slice). The temporal extension is the inverse temperature, $\Nt =1/T$, while the factors $e^{\pm \mu}$ are the lattice implementation of the chemical potential \cite{Hasenfratz:1983ba}. We use the standard notation $\hat 0$ and $\hat i$ for unit steps in the temporal and spatial directions. 
For simplicity we have chosen the staggered signs as $\eta_0=1$ and $\eta_i=(-1)^{x_0+x_1+\ldots+x_{i-1}}$. At zero chemical potential the determinant of the staggered Dirac operator is positive\footnote{The massless Dirac operator anticommutes with $\eta_5$ (the residual chiral symmetry), such that its eigenvalues come in $\pm\lambda$ pairs. Antihermiticity at $\mu=0$ yields eigenvalues on the imaginary 
axis and the positivity of the determinant follows. The mass term only shifts the real part of the 
eigenvalues keeping this positivity.}, but at nonzero real $\mu$ the Dirac operator is no longer antihermitian, as $\dd(\mu)^\dagger=-\dd(-\mu)$, and its determinant is no longer necessarily positive, inducing the sign problem.

In the subset method for QCD we generate subsets by gathering a number of different configurations of the ensemble summing up their individual weights to the subset weight. The members of a subset are generated by multiplying the temporal links $U_0(x)$ with the center $Z_3=\{1,z,z^*\}$ where $z=\exp(2\pi i/3)$. The generated links $U_0$, $z U_0$ and $z^*U_0$  remain in the configuration space, the group SU(3). The invariance of the Haar measure under group multiplications ensures that the integration measure is the same for the three configurations, so that adding the determinants gives the correct weight of the subset (in the strong coupling limit).

\subsection{0+1 dimensions}
\label{sec:oned}

In 0+1 dimensions the configurations are fully characterized by the (untraced) Polyakov line $P$ (which can be shown using a gauge transformation). In Ref.~\cite{Bloch:2013ara} it was shown that the subset $\Omega_P=\{P,zP,z^*P\}$ for one quark flavor has a positive weight for any $P\in \SU(3)$,
\begin{align}
 \sigma(\Omega_P) 
 &= \tfrac13 \left[\det \dd(P)+\det \dd(zP)+\det \dd(z^*P)\right]\nonumber\\
 &= A^3-3A+A|\tr P|^2+2\cosh(3\mu/T)> 0\,, 
\end{align}
where $A=2\cosh\big(\Nt\arsinh\big(\frac{1}{\Nt}\frac{m}{T}\big)\big)\geq 2$, thus solving the sign problem. In the massless case this simplifies to $A=2$ and
\begin{align}
 \sigma(\Omega_P) 
 =2+2|\tr P|^2+2\cosh(3\mu/T)> 0\quad (m=0)\,.
\label{eq:oned_result_massless}
\end{align}
Note that only terms with baryon chemical potential $\mu_b=3\mu$ appear.

\subsection{Higher dimensions}

In higher dimensions each temporal link can be multiplied by an \textit{independent} center element. In other words, the subsets are generated by the direct product of all local $Z_3(x)$,
\begin{align}
 \Omega_{[\mathcal{U}_0,\mathcal{U}_s]}
 &=\{[g\,\mathcal{U}_0,\mathcal{U}_s]|g\in G\}\,,\nonumber\\ 
 G
 &=\bigotimes_{i=1}^V Z_3(x_i)
 \,,\qquad V\equiv \Nt N_1\ldots N_{d-1}\,,
\label{subsets}
\end{align}
or a subgroup thereof. A group element $g=(e^{2\pi i k_1/3},\ldots, e^{2\pi i k_V/3})$ with $k_i\in\{0,1,2\}$ acts on a configuration $\mathcal{U}_0=(U_0(x_1),\ldots,U_0(x_V))$ through 
\begin{align}
 g\,\mathcal{U}_0=(e^{2\pi i k_1/3}U_0(x_1),\ldots,e^{2\pi i k_V/3}U_0(x_V))\,.
\label{eq:group_element}
\end{align}
After introducing subsets the partition function \eqref{ZQCD} can be written as
\begin{align}
 Z&=\int\!d[\mathcal{U}_0,\mathcal{U}_s]\:\sigma(\Omega_{[\mathcal{U}_0,\mathcal{U}_s]})\,,
\end{align}
with subset weights
\begin{align}
 \sigma(\Omega_{[\mathcal{U}_0,\mathcal{U}_s]})
  &=\frac{1}{3^V}\!
  \prod_{i=1}^V\sum_{k_i=0}^2\det\dd([g\,\mathcal{U}_0,\mathcal{U}_s])\,,
  \label{eq:QCD_subsets_partition}
\end{align}
where we just add the determinants because of the invariance of the Haar measure under group multiplication. The cardinality of $G$, $|G|=3^V$, is exponential in the number $V$ of lattice points.

\subsection{Collective subsets}
\label{sec:coll_subsets}

Before discussing the full subsets \eqref{subsets}, let us first consider collective subsets, which contain three configurations. The subset elements are constructed by synchronously rotating the links \emph{on one time slice with the same $Z_3$ element} $\exp(2\pi i k/3)$, $k\in\{0,1,2\}$, and leaving all other links untouched. For the Dirac operator \eqref{eq:Dirac_op} the operation of a collective $Z_3$ rotation can also be interpreted as adding an imaginary chemical potential $\mu/T\to \mu/T +2\pi i k/3$ while keeping the original links $U_0(x)$ unchanged, since the chemical potential can be introduced equivalently through factors $e^{\pm N_t \mu}=e^{\pm \mu/T}$ on one time slice. The partition functions with these shifted chemical potentials are identical, which is the Roberge-Weiss periodicity \cite{Roberge:1986mm}. Nevertheless, for a given configuration the integrands in the three partition functions differ and after adding these up the contributions of the canonical determinants with nonzero 
triality are removed, as we now show.

The fugacity expansion of the partition function, $Z=\sum_q Z_q e^{q\mu/T}$ with canonical partition functions $Z_q$, has a corresponding expansion at the level of the determinants and subset weights,
\begin{align}
 \det \dd([\mathcal{U}_0,\mathcal{U}_s])
 &=\sum_q D_q \, e^{q\mu/T}\,, \nonumber\\
 \sigma(\Omega_{[\mathcal{U}_0,\mathcal{U}_s]})
 &=\sum_q \sigma_q \, e^{q\mu/T}\,,
\label{eq:subsets_canonical}
\end{align}
where the sum ranges over the number of spatial lattice points, $q=-3V_s,\ldots, 3V_s$, with $V_s=N_1\ldots N_{d-1}$, and we have omitted the link arguments of $D_q$ and $\sigma_q$.

As the $Z_3$ rotations can be shifted into an imaginary chemical potential, the collective subsets yield for every $q$ (see also \cite{Bloch:2012bh})
\begin{align}
 \sigma_{q}^{\text{collective}}e^{q\mu/T}=\frac{1}{3}\sum_{k=0}^2 D_q \, e^{q(\mu/T+2\pi i k/3)}\,,
 \label{collective}
\end{align}
and the well-known formula for the sum over powers of roots of unity reduces $q$ to multiples of three,
\begin{align}
  D_q \, e^{q\mu/T}\frac{1}{3}\sum_{k=0}^2 e^{q(2\pi i k/3)}
  &= D_q \, e^{q\mu/T}\delta_{q,3b}\nonumber\\
  \Rightarrow\quad
  \sigma_{q}^{\text{collective}}
  &=D_q\,\delta_{q,3b}
\end{align}
with integer baryon number $b$ in the range $-V_s,\ldots,V_s$.
Therefore, the canonical weights for collective subsets are given by the canonical determinants with \emph{zero triality} and vanish otherwise. It is well-known that the QCD partition function satisfies the same property, i.e.\ only receives triality zero contributions, and can therefore be expanded in the baryon chemical potential $\mu_b=3\mu$. One of the essential mechanisms of subsets is that they ensure this reduction already at the level of the path integrand. Although the collective subsets and the ensuing global reduction to triality zero terms turned out to solve the sign problem in 0+1 dimensions \cite{Bloch:2013ara}, we will see in Sec.\ \ref{sec:num_results} that in higher dimensions they attenuate the sign problem but do not suffice to solve it. Further note that the collective subsets preserve all plaquette values, such that this subset method could be applied directly in full QCD after introducing the gauge action.

For the full subsets \eqref{subsets} the cancellation of the collective subsets is also achieved (since the collective subsets are a subgroup of the full group), but the zero triality weights will be modified further. Full subsets enforce zero triality for each temporal link, i.e.\ even for chemical potentials that would be defined locally, again via $\mu(x)\to  \mu(x)+2\pi ik(x)/3$. This statement will be made more precise in the next section, e.g.\ Eq.~\eqref{eq:mu_triality2}.

\subsection{Subset building blocks}
\label{sec:building_blocks}

In this subsection we will demonstrate how the subset weight can be decomposed into local building blocks. They are the basics for the positivity proof presented in the next section. For that we write the quark determinant in the partition function as an integral over 
Grassmann fields,
\begin{align}
\hspace{-0.25cm}\det \dd(\,\mathcal{U}) = \!\int \left[\prod_{i=1}^V\prod_{a=1}^3 d\psi_a(x_i)d\bar\psi_a(x_i)\right] e^{S_F(\psi,\bar\psi,\,\mathcal{U})},
\end{align}
where $S_F(\psi,\bar\psi,\,\mathcal{U})=\sum_{x,y}\bar\psi(x)D(\mathcal{U};x|y)\psi(y)$ with the staggered Dirac operator of Eq.~\eqref{eq:Dirac_op}. As each term in the fermion action is bilinear in Grassmannians, the exponential can be factorized in single-link contributions
\begin{align}
e^{S_F(\psi,\bar\psi,\,\mathcal{U})} &= \prod_{x,y}\exp\left(\bar\psi(x)D(\mathcal{U};x|y)\psi(y)\right) .
\label{factor}
\end{align}

Let us focus on the contribution of a specific temporal link, i.e.\ $U_0(x)$ and $U_0^\dagger(x)$ for some fixed $x$, which according to \eqref{eq:Dirac_op} is
\begin{align}
 &\ep\bar{\psi}(x)U_0(x)\psi(x+\hat{0})-\emu\bar{\psi}(x+\hat{0})U_0^\dagger(x)\psi(x)\nonumber\\
 &\equiv \ep u + \emu \ub \,,
\label{eq:def_uu'} 
\end{align}
which defines the bosonic variables $u$ and $\ub$.
At the temporal boundary this expression enters the fermionic action with a minus sign. We expand the contribution from this link to the path integral weight
\begin{align}
\exp(\ep u+\emu \ub)
  &=\sum_{n,m=0}^3\frac{1}{n!\,m!}(\ep u)^n (\emu \ub)^m .
\label{eq:expansion_first}  
\end{align}
Because $\psi$ and $\bar\psi$ represent Grassmann fields with three color degrees of freedom, only terms which are at most cubic in $u$ and $\ub$ will contribute.

Center rotations on that link amount to the changes $U_0(x) \to e^{2\pi i k/3} U_0(x)$, and thus $u\to e^{2\pi i k/3}u$ and $\ub\to e^{-2\pi i k/3}\ub$, $k\in\{0,1,2\}$, and the subset sum for a term $(\ep u)^n (\emu \ub)^m$ becomes
\begin{align}
 &\frac{1}{3}\sum_{k=0}^2 (\ep e^{2\pi i k/3}u)^n (\emu e^{-2\pi i k/3}\ub)^m\nonumber\\
  &=(\ep u)^n (\emu \ub)^m
  \frac{1}{3}\sum_{k=0}^2e^{2\pi i k(n-m)/3}
  \label{eq:mu_triality1}\\
  &=\begin{cases} e^{(n-m)\mu}u^n \ub^m & \text{if }(n-m)\text{ mod } 3=0 , 
    \\ 0 & \text{otherwise} . \end{cases}
  \label{eq:mu_triality2}
\end{align}
Terms with nonzero triality are removed, while terms with zero triality remain unchanged. Therefore, the following subset building blocks\footnote{Note, that the symbol $\sigma$ now represents subset sums over Grassmann terms which are still subject to Grassmann integration.} survive from the local contribution \eqref{eq:expansion_first}:
\begin{align}
 \sigma_x \! \equiv \! 1 + u\bar u +\tfrac{1}{2!^2}(u\ub)^2 + \tfrac{1}{3!^2} (u\ub)^3
  +\tfrac{1}{3!} \!\big(\epp u^3+\emum \ub^3\big).
  \label{eq:expansion}
\end{align}
We will denote the $u^3$ term, just three forward hoppings, by `baryonic', the $\ub^3$ term, just three backward hoppings, by `antibaryonic', the terms with $(u\ub)^n$, $n\in\{1,2,3\}$, by `$n$-mesonic' (hoppings) and the identity by `empty link'. Only the (anti)baryonic terms carry the $\mu$-dependence, in the form $e^{\pm 3\mu}=e^{\pm \mu_b}$, with baryon chemical potential $\mu_b$.  

The subset building blocks are similar to those found in the polymer-baryon approach to strong coupling lattice QCD \cite{Rossi:1984cv,Karsch:1988zx}. In the latter the gauge links are integrated out completely, yielding the same hopping structure (in all directions) as in $\sigma_x$. Integrating out the Grassmannians as well, enforces site constraints on the polymers and baryons. In contrast to this, our hopping terms retain a dependence on the background gauge links (actually only the 1- and 2-mesons do, see \eqref{eq:uv_simplification}), but are subject to the same Grassmann constraints.

\section{Positivity}
\label{sec:positivity}

In this section we first present our proof for the positivity on lattices with $\Nt=2$ and arbitrary spatial extent. Then we give a more detailed discussion of the $2\times 2$ lattice, where some concepts become more explicit. While the first subsection uses the framework of Grassmannians, the second subsection makes use of the particular matrix structure of the Dirac operator (which is of course fully equivalent).

\subsection{Positivity proof for $\Nt=2$ using Grassmannians}
\label{sec:ferm_proof}

We first consider the temporal part of the fermionic action and \textit{gather the two points in the temporal direction} at each spatial point, 
\begin{align}
 S_{F,t}=\sum_{\vec{x}}\big(\ep u+\emu\ub-\ep v-\emu \vb\,\big)\,,
 \label{Suv}
\end{align}
where in analogy to \eqref{eq:def_uu'}:
\begin{align}
\begin{aligned}
 u
 &=\,\bar{\psi}(1,\vec{x})U_0(1,\vec{x})\psi(2,\vec{x})\,, \\
 \ub
 &= -\bar{\psi}(2,\vec{x})U_0^\dagger(1,\vec{x})\psi(1,\vec{x})\,,\\
 \label{ubdef}
 v
 &=\,\bar{\psi}(2,\vec{x})U_0(2,\vec{x})\psi(1,\vec{x})\,, \\
 \vb
 &= -\bar{\psi}(1,\vec{x})U_0^\dagger(2,\vec{x})\psi(2,\vec{x})\,,
\end{aligned}
\end{align}
with arguments $\vec{x}$ on the left hand sides omitted for simplicity. The relative minus sign between $u$ and $v$ in \eqref{Suv} comes from the antiperiodic boundary conditions.

For the subset sum we expand $\exp(S_{F,t})$ in analogy to Sec.~\ref{sec:building_blocks}, and again the zero triality condition applies to $(u,\ub)$ and $(v,\vb)$. Using \eqref{eq:expansion} and taking into account the additional sign for $v$ and $\vb$ the contribution of the spatial point $\vec x$ to $\exp(S_{F,t})$ is
\begin{align}
 &\sigma_{(1,\vec{x})}\,\sigma_{(2,\vec{x})} \nonumber\\
 &=
 \left(
  \left[1+\frac{\epp}{3!}\,u^3\right]\left[1+\frac{\emum}{3!}\,\ub^3\right]+u\ub+\frac{(u\ub)^2}{2!^2}
 \right)\nonumber\\
 &\times \big(u\to -v, \ub\to -\vb\big),
\label{eq:another_sigma}
\end{align}
where four of the hopping terms in \eqref{eq:expansion} were collected into a product. 
Because of periodicity, $u$ and $\vb$ visit the same Grassmann variables, although they are connected by different gauge links, and the same holds for $v$ and $\ub$. The Grassmannian antisymmetry restricts \eqref{eq:another_sigma} to polynomials that are cubic both in the combinations $(u,\vb)$ and $(\ub,v)$. Therefore, several cross terms from the product above vanish and one is left with
\begin{align}
 \sigma_{\vec x} \equiv 
 \sigma_{(1,\vec{x})}\,\sigma_{(2,\vec{x})} =
 \sigma_b +\sigma_m\,,
\label{sigma_uv}
\end{align}
where
\begin{align}
 \sigma_b = \left[1+\frac{\epp}{3!}\,u^3-\frac{\emum}{3!}\,\vb^3\right]
 \left[1+\frac{\emum}{3!}\,\ub^3-\frac{\epp}{3!}\,v^3\right]
 \label{eq:sigma_b_def}
\end{align}
combines baryonic terms with the empty link and 3-mesons, and the remainder 
\begin{align}
 \sigma_m
 &=\left(1+u\ub+\frac{(u\ub)^2}{2!}\right)\times \big(u\ub\to v\vb\big)-1\nonumber\\
 &=u\ub+v\vb+\frac{(u\ub)^2}{2!^2}+\frac{(v\vb)^2}{2!^2}+u\ub\, v\vb\nonumber\\
 &+\frac{u\ub(v\vb)^2}{2!^2}+\frac{(u\ub)^2v\vb}{2!^2}
 \label{mesonic_weight}
\end{align}
is a polynomial in $u\ub$ and $v\vb$ and thus mesonic. Note that various contributions in $\sigma$ have Grassmann vacancies and need to combine with spatial hoppings or mass terms to achieve Grassmann saturation for all $\bar{\psi}(x)$ and $\psi(x)$. 

Using the antisymmetry of Grassmannians and $\det U=1$ one can simplify the baryonic terms to be independent of the links%
\footnote{\label{footnote:link_independence}We can write:
\begin{align*}
 \tfrac{1}{3!}u^3 
 &= \tfrac1{3!}(\bar{\psi}U\psiy)^3
  = \tfrac1{3!} (\bar{\psi}_a U_{aa'}\psiy_{a'}) (\bar{\psi}_b U_{bb'}\psiy_{b'}) (\bar{\psi}_c U_{cc'}\psiy_{c'}) \nonumber\\
  &= -\tfrac1{3!}\,(\epsilon_{abc} \epsilon_{a'b'c'} U_{aa'}U_{bb'}U_{cc'})\bar\psi^3\psiy^3\nonumber\\
  &= -\tfrac1{3!}\,(3!\det U) \bar\psi^3\psiy^3
  =-\bar\psi^3\psiy^3,
\end{align*}
where $\psi_a\psi_b\psi_c=\epsilon_{abc}\psi_1\psi_2\psi_3=\epsilon_{abc}\psi^3$ with $\psi^3\equiv\psi_1\psi_2\psi_3$, and we used the Leibniz formula for the determinant and ${\det U = 1}$. The minus signs comes from the permutation of Grassmann variables.} 
\begin{align}
\begin{aligned}
 \frac{u^3}{3!} = -\frac{\vb^3}{3!} 
 &= - \bar{\psi}^3(1,\vec{x})\psi^3(2,\vec{x})\,,\\ 
 \frac{v^3}{3!} = -\frac{\ub^3}{3!} 
 &= - \bar{\psi}^3(2,\vec{x})\psi^3(1,\vec{x})\,, 
 \label{eq:uv_simplification}
\end{aligned}
\end{align}
where $\psi^3 \equiv \psi_1\psi_2\psi_3$. 
The baryonic product can therefore be written as
\begin{align}
 \sigma_b 
 &=\left[1+2\cosh(3\mu)\frac{u^3}{3!}\right]
   \left[1+2\cosh(3\mu)\frac{\ub^3}{3!}\right] 
 \notag\\
 &=\left[1-2\cosh(3\mu)\bar{\psi}^3(1,\vec{x})\psi^3(2,\vec{x})\right]\nonumber\\
 &\times\left[1+2\cosh(3\mu)\bar{\psi}^3(2,\vec{x})\psi^3(1,\vec{x})\right] .
   \label{eq:cosh_cosh}
\end{align}

With \eqref{eq:cosh_cosh} and \eqref{mesonic_weight} the subset sum \eqref{sigma_uv} for the temporal links at each site $\vec x$ can be written as
\begin{align}
 \sigma_{\vec x} 
 &=\left[1-2\cosh(3\mu)\bar{\psi}^3(1,\vec{x})\psi^3(2,\vec{x})\right]\nonumber\\
 &\cdot
   \left[1-2\cosh(3\mu)\psi^3(1,\vec{x})\bar{\psi}^3(2,\vec{x})\right] 
   \notag\\
 &+u\cdot \ub+\vb\cdot v
   +\frac{u^2}{2}\cdot\frac{\ub^2}{2}
   +\frac{\vb^2}{2}\cdot\frac{v^2}{2}+u\vb\cdot\ub v\nonumber\\
 &+\frac{u\vb^2}{2}\cdot\frac{\ub v^2}{2}
   +\frac{u^2\vb}{2}\cdot\frac{\ub^2 v}{2} ,
   \label{eq:sigma_rewritten}
\end{align}
where we have rewritten the terms such that the first factor depends on $\{\bar{\psi}(1,\vec{x}), \psi(2,\vec{x})\}$ and the second on $\{\psi(1,\vec{x}), \bar{\psi}(2,\vec{x})\}$, cf.\ the definitions \eqref{ubdef} of $u,\ub,v$ and $\vb$ (for clarity the multiplication of such factors is represented by a dot). With appropriate functions $h_\alpha$ the subset weight \eqref{eq:sigma_rewritten} can be summarized as
\begin{align}
 \sigma_{\vec x} 
 = \sum_{\alpha=0}^7\:
 & h_\alpha(\bar{\psi}(1,\vec{x}), \psi(2,\vec{x}),U_0(1,\vec{x}),U_0(2,\vec{x})^*)\nonumber\\
 \cdot &\, h_\alpha(\psi(1,\vec{x}), \bar{\psi}(2,\vec{x}),U_0(1,\vec{x})^*,U_0(2,\vec{x})) .
 \label{halpha}
\end{align}
In each term both factors can be obtained from one another by exchanging the fermions and antifermions, $\psi\rightleftharpoons\bar{\psi}$, and complex conjugating the links. After denoting this exchange operation as
\begin{align}
 \Gconj{f}\equiv f\big|_{\psi\rightleftharpoons\bar{\psi},\,U_\nu\rightleftharpoons U_\nu^*}\,,
 \label{Gconj}
\end{align}
we can write
\begin{align}
 \sigma_{\vec x} =\sum_{\alpha} h_\alpha(\vec{x})\cdot\Gconj{h_\alpha(\vec{x})} .
 \label{halpha1}
\end{align}
For the spatial part of the fermion action, see \eqref{factor}, we find, after anticommuting the fermions in the backward hopping term,
\begin{align}
 S_{F,s}
 &=\sum_{x}\sum_{i} \left(\bar{\psi}_a(x)\eta_i(x)U_{i,ab}(x)\psi_b(x+\hat{i})\right.\nonumber\\
 &\qquad\qquad\left.+\,\psi(x)_a\eta_i(x)U_{i,ab}^*(x)\bar{\psi}_b(x+\hat{i})\right)
 \label{eq:spat_hopping_modified}\\
 &\equiv \sum_{x} \left(w(x)+\Gconj{w(x)}\right) ,
\end{align}
where we explicitly wrote out the color indices in \eqref{eq:spat_hopping_modified} (with implicit summation over repeated indices) to easily identify the exchange symmetry \eqref{Gconj}. The spatial weight (without subsets for the spatial links) is then
\begin{align}
 e^{S_{F,s}}
 &=\prod_{x}\exp\big(w(x)\big)\cdot\Gconj{\exp\big(w(x)\big)} .
 \label{Sspatial}
\end{align}
This embodies the antihermiticity of the Dirac operator, which is an important ingredient in the positivity of the determinant at $\mu=0$, and hints at the fact that the structure \eqref{halpha1} will be important to prove the positivity of the subset weights.

\newcommand{\A}{A}

The second ingredient necessary to show the positivity of the subsets is the staggered chirality, i.e.\ the fact that all interactions connect even with odd sites or odd with even sites (recall that $m=0$). Here even and odd lattice sites $x^e$ and $x^o$ are those with $\eta_5(x)=+1$ and $-1$, respectively, where $\eta_5=(-1)^{x_0+x_1+\cdots+x_{d-1}}$. 
In all products $h_\alpha\cdot\Gconj{h_\alpha}$ in \eqref{halpha1} and $e^w\cdot\Gconj{e^w}$ in \eqref{Sspatial} one factor will only depend on the Grassmann sets $\{\bar\psi(x^e)\}\equiv \bar\Psi^e$ and $\{\psi(x^o)\}\equiv\Psi^o$ coming with links $\{U_\nu(x^e)\}\equiv{\cal U}^e$ and $\{U_\nu^*(x^o)\}\equiv{\cal U}^{o*}$ and the other only on the complements $\Psi^e$, $\bar\Psi^o$ with links ${\cal U}^{e*}$, ${\cal U}^{o}$. The full massless subset weight reads
\begin{align}
 \sigma
 &=
 \prod_{\vec{x}} \sum_{\alpha_{\vec{x}}} h_{\alpha_{\vec{x}}}(\vec{x})\cdot\Gconj{h_{\alpha_{\vec{x}}}(\vec{x})}
\; \prod_{x}\exp(w(x))\cdot\Gconj{\exp(w(x))},
\label{sigma_4.17}
\end{align}
and after expanding all products we can rewrite $\sigma$ as
\begin{align}
 \sigma
 \!&=\! \sum_{\A}
 f_{\A}\left(\bar\Psi^e,\Psi^o,{\cal U}^e,{\cal U}^{o*}\right)\,
 \!\cdot\! f_{\A}\left(\Psi^e,\bar\Psi^o,{\cal U}^{e*},{\cal U}^{o}\right) ,
\end{align}
where the index $\A=\{\alpha_{\vec{x}}\}$ runs over all combinations of $\alpha_{\vec{x}}$ for all spatial points $\vec{x}$ and oversaturated terms, i.e. with too many Grassmannians, automatically cancel.

Moreover, the Grassmann measure factorizes analogously
\begin{align}
 &\prod_x d\psi(x) d\bar{\psi}(x)
= \prod_{x^e}\prod_{x^o} d\psi(x^e) d\bar{\psi}(x^e) d\psi(x^o) d\bar{\psi}(x^o)\nonumber \\
 &\quad= \prod_{x^e} d\bar\psi(x^e) \prod_{x^o} d\psi(x^o)
 \cdot \prod_{x^e}  d\psi(x^e) \prod_{x^o} d\bar\psi(x^o)\nonumber\\
 &\quad\equiv d\bar\Psi^e d\Psi^o \cdot d\Psi^e d\bar\Psi^o\,, 
\end{align}
where $d\bar\psi d\psi = \prod_{a=1}^3 d\bar\psi_a d\psi_a$. The Grassmann integrals over $\sigma$ remove all terms with insufficient Grassmann content and each of the surviving terms is a product of two polynomials in the (complex) links only. By renaming integration variables it can be seen that these polynomials are still related by complex conjugation of the link arguments
\begin{align}
 \!\!\int\prod_x d\psi(x) d\bar{\psi}(x)\, \sigma\!
 &=\sum_{\A}
 \int d\bar\Psi^e d\Psi^o 
 f_{\A}\left(\bar\Psi^e,\Psi^o,{\cal U}^e,{\cal U}^{o*}\right)\nonumber\\
 &\cdot\int d\Psi^e d\bar\Psi^o
 f_{\A}\left(\Psi^e,\bar\Psi^o,{\cal U}^{e*},{\cal U}^{o}\right)\nonumber\\
 &=\!\sum_{\A}
 p_{\A}\left({\cal U}^{e},{\cal U}^{o*}\right)
 p_{\A}\left({\cal U}^{e*},{\cal U}^{o}\right).\!
\end{align}
Since the polynomials $p$ emerge from the hopping terms in the Dirac operator, they possess \textit{real coefficients} only. Therefore, polynomials with complex conjugated link arguments turn into complex conjugated polynomials, such that
\begin{align}
 \int\prod_x d\psi(x) d\bar{\psi}(x)\, \sigma
 &=\sum_{\A}
 p_{\A}\left({\cal U}^e,{\cal U}^{o*}\right)
 \big[p_{\A}\left({\cal U}^e,{\cal U}^{o*}\right)\big]^*\nonumber\\
 &=\sum_{\A}
 \big|p_{\A}\left({\cal U}^e,{\cal U}^{o*}\right)\big|^2\,
 \geq 0 .
\end{align}
This proves that the massless subset weight is positive for $N_t=2$ at nonzero chemical potential and arbitrary spatial extent. What is more, the subset weight consists of various positive subterms (labeled by $A$).

The modification of this proof caused by a nonzero mass is presented in Appendix \ref{app:massive}. For $\mu=0$ the subset construction is not needed and one can prove the positivity by using \eqref{Sspatial} for all directions.

In showing the positivity we have made use of the fact that for $\Nt=2$, $u$ and $\vb$ connect the same sites and thus have the same Grassmann content but with opposite $\mu$-dependence, and that the baryonic factors are independent of the connecting links, \eqref{eq:uv_simplification}, such that $\mu$ enters in the form \eqref{eq:cosh_cosh}. This would not hold with nonzero triality terms present, as is the case in the determinant formulation without subsets.

Let us make some further remarks. Firstly, note that the temporal antiperiodicity is crucial to ensure the subset positivity. Consider for instance periodic boundary conditions. In this case the $v$- and $\vb$-terms in \eqref{Suv}, \eqref{eq:another_sigma} and \eqref{eq:sigma_b_def} would have plus signs instead of minus signs, and all odd powers of $v$ and $\vb$ in the derivation above would have opposite signs. The mesonic discussion is left unchanged as $v$ and $\vb$ always come in pairs, but the baryonic product would become
\begin{align}
 &\left[1+\frac{\epp}{3!}u^3+\frac{\emum}{3!}\vb^3\right]
 \left[1+\frac{\emum}{3!}\ub^3+\frac{\epp}{3!}v^3\right] \nonumber\\
 &= \left[1+2\sinh(3\mu)\frac{u^3}{3!}\right]\left[1-2\sinh(3\mu)\frac{\ub^3}{3!}\right]\,.
\end{align}
 Due to the different signs in both factors, this product cannot be written as $h\cdot\Gconj{h}$ and no longer satisfies the conjugation symmetry required in the positivity proof above.

Note that the last two terms of the mesonic weight \eqref{mesonic_weight} can be simplified further as (see Appendix \ref{app:Meson-contrib})
\begin{align}
 &u\ub (v\vb)^2 
 = (u\ub)^2 v\vb\nonumber\\
 &= 4 \, |\tr P(\vec{x})|^2
 \bar{\psi}^3(1,\vec{x})\psi^3(1,\vec{x}) \bar{\psi}^3(2,\vec{x})\psi^3(2,\vec{x})\,.
\label{eq:simplify_to_Ploops}
\end{align}
This and the $4\cosh^2(3\mu)$-term from \eqref{eq:cosh_cosh}, which equals $2+2\cosh(6\mu)=2+2\cosh(3\mu/T)$,
have full Grassmann content and represent the full weight for the one-dimensional massless case at $\Nt=2$, cf.\ \eqref{eq:oned_result_massless}.

The last remark is slightly more formal. One can rewrite the subset contribution $\sigma$ as an exponential of an effective subset action,
\begin{align}
 \sigma
 &\propto \exp\sum_{\vec{x}}\Big( u\ub+v\vb-\frac{(u\ub)^2+(v\vb)^2}{4}\nonumber\\
 &-6\,\bar{\psi}^3(1,\vec{x})\psi^3(1,\vec{x}) \bar{\psi}^3(2,\vec{x})\psi^3(2,\vec{x})
 \label{subsetS}\\
 & +2\cosh(3\mu)\big\{\bar{\psi}^3(1,\vec{x})\psi^3(2,\vec{x})
     -\bar{\psi}^3(2,\vec{x})\psi^3(1,\vec{x})\big\}\Big)\,,\nonumber
\end{align}
which can easily be checked by an expansion of the exponential function (which again terminates).
In contrast to the Dirac action this effective action is not bilinear in $(\bar{\psi},\psi)$ but also involves higher powers of the Grassmann fields. Therefore the Grassmann integral of $\sigma$ can not be represented as a determinant.

\subsection{Alternative proof for the massless $2\times2$ lattice}
\label{sec:more_explicit}

Below we give an alternative positivity proof, which only holds for a $2\times2$ lattice in the massless case. The salient feature of this proof is that it directly uses the determinant formulation.

For a $2\times 2$ lattice the Dirac operator can be written as 
\begin{align}
 \dd = 
 \begin{pmatrix}
   0 & 0 & T^{\eo}_1 & S_1 \\
   0 & 0 & S_2 & T^{\eo}_2 \\
   -T^{\oe}_1 & -S_2^\dagger & 0 & 0\\
   -S_1^\dagger & -T^{\oe}_2 & 0 & 0
 \end{pmatrix},
\label{eq:Dirac_op_2x2} 
\end{align}
where $T_x$ and $S_t$ are the temporal and spatial hoppings on the spatial slice $x$ and time slice $t$, respectively. These $3\times3$ blocks are given by
\begin{align}
\begin{aligned}
 T^{\eo}_1 &= \ep U_0(11)+\emu U_0^\dagger(21)\,,\\
 T^{\oe}_1 &= \ep U_0(21)+\emu U_0^\dagger(11)\,,\\
 T^{\eo}_2 &= -\Big[\ep U_0(22)+\emu U_0^\dagger(12)\Big]\,,\\
 T^{\oe}_2 &= -\Big[\ep U_0(12)+\emu U_0^\dagger(22)\Big]\,,
 \label{eq:T_one}
 \\
 S_1 &= -\Big[U_1(11)-U_1^\dagger(12)\Big]\,,\\
 S_2 &= U_1(22)-U_1^\dagger(21)\,,
\end{aligned}
\end{align}
where the superscripts $\eo$ and $\oe$ stand for even-odd and odd-even hoppings. Each entry is a sum of two contributions because neighboring sites on a $2\times2$ lattice can be connected in two ways, where one is `around the world'. The spatial part of the Dirac operator is antihermitian, therefore only two independent $S_t$ occur. At $\mu=0$ the full Dirac operator is antihermitian and $T^{\oe}_x=(T^{\eo}_x)^\dagger$. The sign difference between $S_1$ and $S_2$ is due to the staggered phase, whereas the signs in $T_x$ reflect the antiperiodic boundary conditions.

Using the determinant formula for block matrices we find
\begin{align}
 \det \dd 
 &= \det \begin{pmatrix}
   T^{\eo}_1 & S_1 \\
   S_2 & T^{\eo}_2 
 \end{pmatrix}
 \det \begin{pmatrix}
   -T^{\oe}_1 & -S_2^\dagger\\
   -S_1^\dagger & -T^{\oe}_2
 \end{pmatrix}\nonumber\\
 &= \det(SS^\dagger) \det(\one_3-M^{\eo})\det(\one_3-M^{\oe})\,,
 \label{SMM}
\end{align}
where we defined
\begin{align}
\begin{aligned}
 S 
 &= S_1 S_2 \,, \\
 M^{\eo} 
 &= T^{\eo}_1 S_2^{-1} T^{\eo}_2 S_1^{-1} \label{eq:X_def},\\
 M^{\oe} 
 &= [S_1^{-1}]^\dagger T^{\oe}_2[S_2^{-1}]^\dagger T^{\oe}_1 \,. 
\end{aligned} 
\end{align}
The first factor of \eqref{SMM} is the determinant of the spatial part of the Dirac operator, which is blind to $\mu$ and therefore positive. Formally, the positivity follows because a matrix product $S S^\dagger$ is always positive-semidefinite. At $\mu=0$ the $M$-matrices are related as $M^{\oe}=(M^{\eo})^\dagger$, and the full determinant is positive for the same reason.

The last two determinants of $3\times 3$ matrices in \eqref{SMM} can be expanded as 
\begin{align}
 \det(\one_3-M)\stackrel{3\times 3}{=} 1-\det M -\tr M+\frac{(\tr M)^2-\tr M^2}{2} ,
\end{align}
which can, for example, be proven in terms of the eigenvalues of $M$. 
Before constructing the full subsets, we first construct `coarse subsets' $\Omega_c$ containing three configurations by multiplying $U_0(11)$ with the three center phases and $U_0(21)$ by their complex conjugate. The remaining links are left untouched. These rotations form a subgroup of the full subset group. They multiply $T_1^{\eo}$ and thus $M^{\eo}$ by the three center phases and $T_1^{\oe}$ and thus $M^{\oe}$ by the complex conjugate phases, and consequently
\begin{align}
\begin{aligned}
 \det(\one_3-e^{2\pi i k/3} M^{\eo})
  &=1-\det M^{\eo} -e^{2\pi i k/3} \tr M^{\eo}\\
  &+e^{4\pi i k/3} \frac{(\tr M^{\eo})^2-\tr M^{\eo\,2}}{2}\, ,\\
  \!\!\!\!\!\det(\one_3-e^{-2\pi i k/3} M^{\oe})
  &=1-\det M^{\oe} -e^{-2\pi i k/3} \tr M^{\oe}\\
  &+e^{-4\pi i k/3} \frac{(\tr M^{\oe})^2-\tr M^{\oe\,2}}{2}\, .
\end{aligned}  
\end{align}
The coarse subset sum of the product \eqref{SMM} can then be computed as in \eqref{eq:mu_triality2} and we find
\begin{align}
&\tfrac13\sum_{\Omega_c} \det D =
 \det(SS^\dagger) \, \Bigg[(1-\det M^{\eo})(1-\det M^{\oe}) \nonumber \\
&\!+\tr M^{\eo}\tr M^{\oe} 
+\frac{(\tr M^{\eo})^2-\tr M^{\eo\,2}}{2}
  \frac{(\tr M^{\oe})^2-\tr M^{\oe\,2}}{2}\Bigg] . \label{coarse}
\end{align}

After completing the full subsets the chemical potential cancels in the second line of \eqref{coarse}, since it can only enter if a temporal link or its inverse appears three times, and by inspection of \eqref{eq:T_one} and \eqref{eq:X_def} this cannot happen. Since this expression is independent of $\mu$, its value can equally well be computed at $\mu=0$ where $M^\oe=(M^\eo)^\dagger$. This gives the following full subset contribution for this second line
\begin{align}
\sigma_{(II)} 
&= |\det S|^2 \sigma_{0}\Big( \big|\tr M^{\eo}\big|^2 
 + \tfrac14\big|(\tr M^{\eo})^2-\tr M^{\eo\,2}\big|^2 \Big)\label{sigmaII}
\end{align}
where $\sigma_0(\cdots)$ indicates the subset at $\mu=0$, and clearly $\sigma_{(II)}$ is positive. Full subsets of the first line of \eqref{coarse} give a contribution
\begin{align}
 \sigma_{(I)} 
 &= \frac1{|\Omega|}\sum_\Omega \det (SS^\dagger) \, \Big[1-\det(S^{-1})\det T^{\eo}_1\det T^{\eo}_2\Big]\nonumber\\
 &\times\Big[1-\det(S^{-1}\big)^\dagger\det T^{\oe}_1\det T^{\oe}_2\Big] ,
 \label{sigmaI}
\end{align}
where we substituted $M^{\eo,\oe}$, defined in \eqref{eq:X_def}. The subset sums for temporal hoppings are computed in Appendix\ \ref{app:temp_hoppings} and substitution of \eqref{eq:det_subsetted} yields,
 \begin{align}
 \sigma_{(I)} 
 &= |\det S|^2 + (\det S + \det S^\dagger)\, 4 \cosh^2(3\mu)\nonumber\\
 & +4\big(2\cosh^2(3\mu)+|\tr P(1)|^2\big)\big(P(1)\to P(2)\big) \nonumber\\
 &= \left| 4 \cosh^2(3\mu) + \det S\right|^2\nonumber\\
 & + 8 \cosh^2(3\mu) \big(|\tr P(1)|^2+|\tr P(2)|^2\big) \nonumber\\
 & + 4 \,|\tr P(1)|^2|\tr P(2)|^2 ,
 \label{sigmab}
\end{align}
where the Polyakov loops $P$ only depend on the spatial argument after tracing. This expression contains mesonic ($\mu$-independent) and baryonic ($3\mu$-dependent) terms.

The full subset weight is simply 
\begin{align}
 \sigma_{\Omega_{[\mathcal{U}_0,\mathcal{U}_s]}} &= \sigma_{(I)} + \sigma_{(II)} \label{sigmatot} \,,
 \end{align}
with $\sigma_{(I)}$ and $\sigma_{(II)}$ given in \eqref{sigmab} and \eqref{sigmaII}.
As all the summands are positive the subset weight is positive too.  It is interesting to note that the first term explicitly combines a $\mu$-independent and a $\mu$-dependent contribution to achieve its positivity. A similar principle is at work in the first term of \eqref{eq:sigma_rewritten} in the more general proof given in Sec.\ \ref{sec:ferm_proof}.

Finally, let us have a look at the canonical subset weights (integrands of the subsets canonical partition functions) as introduced in Eq.~\eqref{eq:subsets_canonical}. This means nothing but collecting the terms in the subset result according to their $\mu$-factors. The terms $e^{\pm 12 \mu}=e^{\pm 6\mu/T}$ come from the  term $16\cosh^4(3\mu)=(e^{3\mu}+e^{-3\mu})^4$ in \eqref{sigmatot} and have weights $\sigma_{q=\pm 6}=1$ as expected from (anti)baryon saturation of the lattice. The weights of the next terms $e^{\pm 6 \mu}=e^{\pm 3\mu/T}$ are
\begin{align}
 \sigma_{q=\pm 3}=\det S + (\det S)^* +4+2|\tr P_1|^2+2|\tr P_2|^2 .
 \label{sigma3}
\end{align}
To compute the first two terms we note that
\begin{align}
 \det S_2
 &=\det\big(\underbrace{U_1(22)U_1(21)}_{\textstyle W(2)}-1\big)
  \underbrace{\det U_1^\dagger(21)}_{\textstyle 1}\nonumber \\
 &=\tr W(2)-\tr W(2)^\dagger
 =2i\,\text{Im}\tr W(2)\,,
 \end{align}
where $W(t)$ is the Wilson loop on time slice $t$ closing around the spatial boundaries, and the second equality can easily be checked for any SU(3) matrix in term of its eigenvalues. Similarly we find for the other time slice
 \begin{align}
 \det S_1 &=-2i\,\text{Im}\tr W(1) ,
\end{align}
where the minus sign comes from the staggered phase. On multiplying we find
 \begin{align}
 \det S\phantom{_1}
 &=4 \,\text{Im}\tr W(1) \,\text{Im}\tr W(2)\,.
 \label{detS}
\end{align}
The first two terms of \eqref{sigma3} can thus be negative and are not necessarily compensated for by the remaining positive terms in $\sigma_{q=\pm 3}$. Therefore, the subset positivity does not generically hold for the individual \textit{canonical} subset weights.

\subsection{Note on spatial subsets}

In this work we apply center subsets to temporal links only, which we conjecture are the minimal subsets for achieving positivity of subset weights. The main motivation for this choice was that the chemical potential causing the sign problem only couples to the temporal hoppings. Extending the idea and applying center subsets on the spatial links as well is another option to compute the partition function, which we briefly comment on. 

As the temporal subsets are positive already, the spatial subsets just add positive numbers and thus remain positive.
Obviously, the cost for such subsets is even bigger, namely $3^{Vd}$, which is why we have not used spatial subsets in practice.

The usefulness of spatial subsets can be seen at the level of the canonical subsets, which are not necessarily positive as  discussed at the end of the previous section. Consider the term \eqref{detS} for the $2\times2$ lattice, which can cause a negative contribution to $\sigma_{q=\pm 3}$ in \eqref{sigma3}. This term is linear in the Wilson loops $W(1)$ and $W(2)$ and will disappear after spatial subsetting according to Eq.~\eqref{eq:mu_triality2} (with $(n,m)=(1,0)$ or $(0,1)$). In this particular case even the canonical subset weight becomes positive upon spatial subsetting. 

In future work we plan to use the positive summands obtained after subsetting to sample the partition function with a worm algorithm. In this context the spatial subsets could give an additional advantage as they further reduce the number of allowed building blocks, while still keeping the dependence on the SU(3) links.

\section{Numerical results}
\label{sec:num_results}

We implemented the full subset method in a Monte Carlo simulation. The subset weights are computed explicitly by adding the numerically computed determinants for all the configurations belonging to the subset. These positive subset weights are then used to generate relevant subsets of the partition function using a Metropolis importance sampling algorithm.

\begin{table*}
\begin{center}
\small
\begin{tabular}{|c|c|l|l|l|l|l|}
\hline
& $N_t\times N_x$ & 
\multicolumn{1}{c|}{$2\times2$} &
\multicolumn{1}{c|}{$4\times2$} &
\multicolumn{1}{c|}{$6\times2$} &
\multicolumn{1}{c|}{$8\times2$} &
\multicolumn{1}{c|}{$10\times2$} 
\\
\hline
a & phase-quenched & 0.8134(3) & 0.4361(4) & 0.233(2) & 0.130(2) & 0.071(1) \\
b & sign-quenched & 0.9271(2) & 0.6150(5) & 0.355(3) & 0.203(2) & 0.109(2) \\
c & collective & 0.9778(9) & 0.777(4) & 0.500(6) & 0.303(8) & 0.178(3) \\
d & T-slice & 1.0 & 0.9896(5) & 0.885(2) & 0.670(5) & 0.436(8) \\
e & full & \textbf{1.0}   & \textbf{1.0}  & \textbf{1.0}$^*$ & \textbf{1.0}$^*$ & \textbf{1.0}$^*$ \\
\hline
\end{tabular}
\hspace*{6.5mm}\begin{tabular}{|c|c|l|l|l||l||l|}
\hline
& $N_t\times N_x$ & 
\multicolumn{1}{c|}{$2\times 4$} &
\multicolumn{1}{c|}{$4\times 4$} &
\multicolumn{1}{c||}{$6\times 4$} &
\multicolumn{1}{c||}{$2\times 6$} &
\multicolumn{1}{c|}{$2\times 8$} 
\\
\hline
a & phase-quenched & 0.7934(5) &  0.295(1) & 0.0961(9) & 0.7364(6) & 0.6725(7) \\
b & sign-quenched & 0.9197(3) & 0.442(2)& 0.149(1)  & 0.8917(4)  & 0.8523(5) \\
c & collective & 0.959(1) & 0.557(6)& 0.214(8) & 0.912(3) & 0.867(2) 	\\
d & T-slice & 1.0 & 0.9973(2) & 0.812(3) & 1.0 & 1.0 \\
e & full & \textbf{1.0}$^*$ & \textbf{1.0}$^{*}$ & \textbf{1.0}$^{*}$ & \textbf{1.0}$^{*}$  & \textbf{1.0}$^{*}$ \\
\hline
\end{tabular}
\caption{Reweighting factors for 2d-QCD for $N_f=1$ ($m=0$) for (a) phase-quenched and (b) sign-quenched reweighting in the link-formulation, and for (c) collective subsets, (d) T-slice subsets and (e) full subsets. The columns give the data for $\Nt\times2$ grids with $\Nt=2,4,6,8,10$,  for a $\Nt\times4$ grid with $\Nt=2,4,6$ and for a $2\times6$ and $2\times8$ grid, all in the strong-coupling limit at $\mu=0.3$ with $\Nmc=100,000$ ($^*$ means $\Nmc=1,000$).\vspace{-2ex}}
\label{Tab:rewfac}
\end{center}
\end{table*}

As the full subset contains an exponential number of configurations we speed up the computation in a number of ways:
\begin{itemize}
\item There is some redundancy in the full subset: although by definition the full subsets contain $3^{V_s\Nt}$ configurations, there is a $3^{\Nt-1}$ fold degeneracy in the determinant values so that we effectively only need to consider $3^{(V_s-1)\Nt+1}$ different configurations per subset.

This degeneracy occurs when all temporal links inside time slice $i$ are rotated by the same $Z_3$-factor $z_i$, $i=1,\ldots,\Nt$, while the Polyakov lines are left unchanged, i.e.\ $\prod_{i=1}^{\Nt} z_i = 1$. These constrained rotations lead to a $3^{\Nt-1}$ degeneracy of each determinant value. This equality of determinants can be understood as the latter consist of closed loops and the rotations leave all loop values unchanged: for loops involving temporal links we either encounter $z_i$ and $z_i^*$ if the loop goes backward and forward in time, albeit at different spatial points, or we wrap around the lattice which also yields unity because of the constraint.

\item Rank-6 corrections are used to reduce the numerical work in the computation of the determinants when stepping from one  configuration to the next in the subset.

\item The algorithm is efficiently parallelized by evenly distributing the configurations of each subset over several threads.
\end{itemize}
More details on the numerical implementation will be given in a forthcoming publication.

\begin{figure*}
\framebox{
\includegraphics[width=0.8\textwidth]{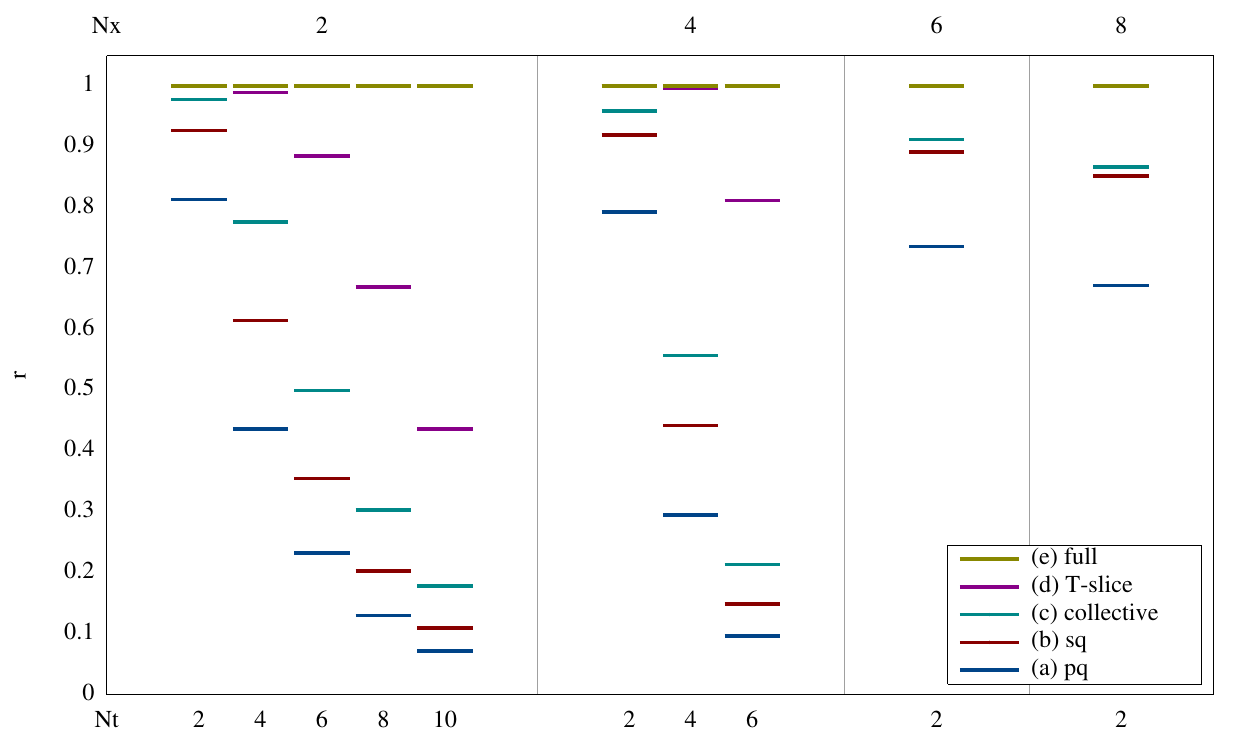}
}
\caption{Reweighting factors $r$ for 2d-QCD for $N_f=1$ ($m=0$) in the strong coupling limit at $\mu=0.3$, where the sign problem is largest, with $\Nmc=100,000$ (for the larger lattices the full subsets were simulated with $\Nmc=1,000$, see Table \ref{Tab:rewfac}). We compare the reweighting factors in the link-formulation for phase-quenched (pq) and sign-quenched (sq) reweighting, and in the subset formulation for collective, T-slice and full subsets. Each column shows the different (color coded) reweighting factors for a specific $\Nt\times N_x$ lattice. For $\Nt=2$ the T-slice data are not visible as they overlap with the full direct product data ($r=1.0$ for both of them).}
\label{Fig:rewfac}
\end{figure*}

In Table \ref{Tab:rewfac} and Fig.\ \ref{Fig:rewfac} we present results obtained with direct product subsets for QCD in two dimensions with massless staggered quarks.
We compare the average reweighting factors for two conventional reweighting schemes in the link formulation:

\begin{enumerate}[label=(\alph*),leftmargin=7mm]
 \item phase-quenched reweighting, and

 \item sign-quenched reweighting,
\end{enumerate}
with the reweighting factors for three different subset constructions:
\begin{enumerate}[label=(\alph*),leftmargin=7mm,start=3]
 \item a collective subset constructed by synchronous $Z_3$ rotations of all temporal links on one time slice, as  mentioned in Sec.~\ref{sec:coll_subsets},

 \item a direct product of local $Z_3$ subsets for the temporal links of all spatial sites \emph{on one time slice}  (containing $3^{V_s}$ configurations), which we call T-slice subsets, and 

 \item the full subsets \eqref{subsets}, which is a direct product of local $Z_3$ subsets for the temporal links on \emph{all} lattice sites.
\end{enumerate}

For the full subsets data were collected for $\Nt\times2$ grids with $\Nt=2,4,6,8,10$, for $\Nt\times4$ grids with $\Nt=2,4,6$ and for a $2\times6$ and $2\times8$ grid, all for $N_f=1$ and $m=0$ in the strong-coupling limit. As can be seen from the phase-quenched reweighting factor (a) the sign problem steadily grows as $\Nt$ and $N_x$ is increased. The sign quenched reweighting (b) somewhat reduces the sign problem, which can be useful for simulations at small chemical potential \cite{deForcrand:2002pa}. Whereas a collective $Z_3$ rotation (c) does not bring much improvement in the two-dimensional case, the T-slice subsets (d) substantially improves on the sign problem. However, the truly surprising observation is that the full subsets (e) yield subset weights that are real\footnote{The imaginary part of the weights is trivially cancelled by implicitly pairing each configuration with its complex conjugate.} and positive in all cases considered. We have proven this property for $\Nt=2$ in Sec.~\ref{sec:positivity}, but conjecture that it holds for any lattice size in any dimension.

\begin{figure}
\includegraphics[width=0.5\textwidth]{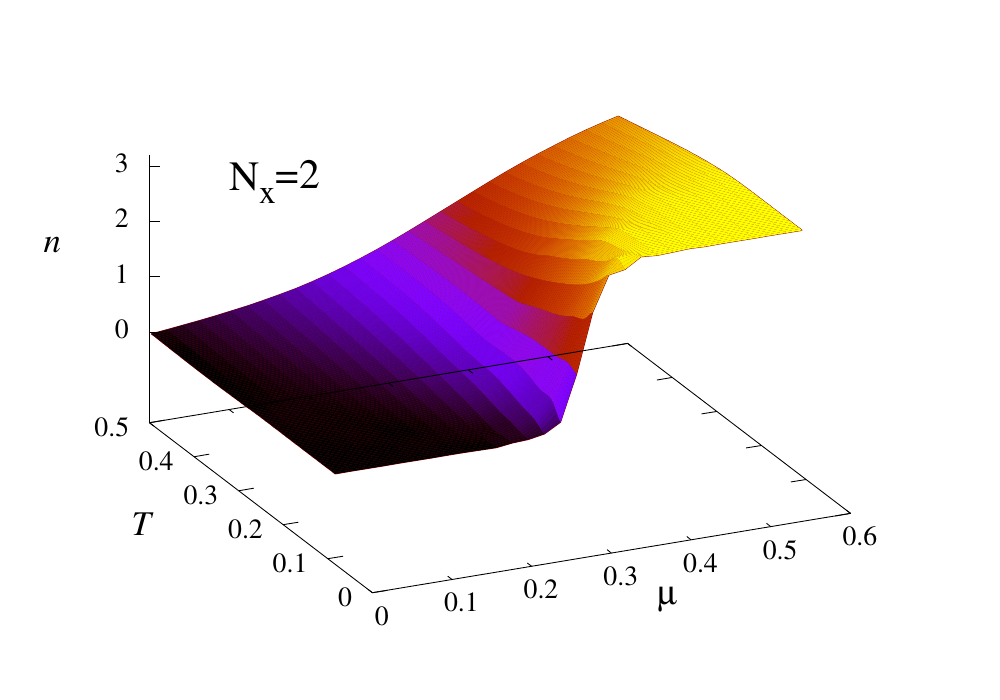}
\vspace{-5mm}
\includegraphics[width=0.5\textwidth]{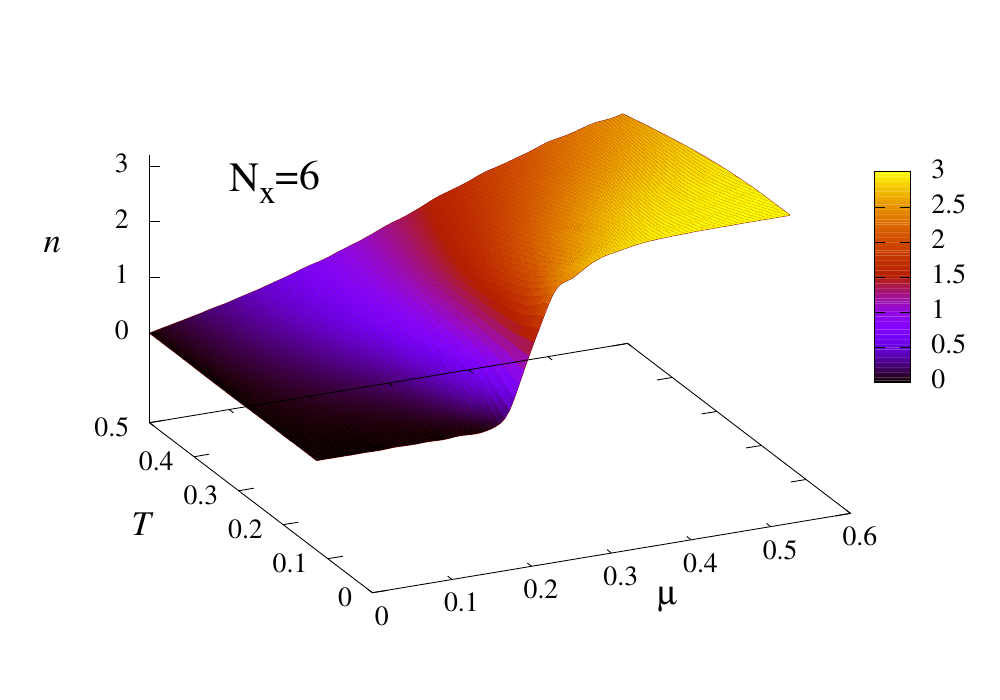}
\vspace{-3mm}
\caption{Surface plots of the quark number density $n$ versus temperature $T$ and quark chemical potential $\mu$ (in lattice units) for $N_x=2$ ($N_t=2\ldots12$) and $N_x=6$ ($N_t=2\ldots8$).\vspace{-1mm}}
\label{Fig:qn_3dplot}
\end{figure}

As an application of the subset method we show the quark number density as a function of $T$ and $\mu$ (in lattice units) in Fig.~\ref{Fig:qn_3dplot} for QCD in 1+1 dimensions (for the computation of observables in the subset framework, see \cite[Eq.\ (3.4)]{Bloch:2013ara}). One observes the Silver Blaze phenomenon where the quark number is independent of $\mu$ below some value $\mu_c$ when $T\to 0$. For the larger lattices the full subsets (e) were too costly and we used T-slice subsets (d) instead. These subsets require additional reweighting (away from $N_t=2$), but still yield a vast improvement over the standard phase quenched reweighting method.

We also verified the effect of the gauge action on the reweighting factors for the $2\times6$ lattice by switching on $\beta$ to leave the strong-coupling regime. The subset weights have to be modified to take into account the different values of the gauge action for the different subset elements, and the sign problem slowly reappears even for the full product subsets. Nevertheless, for $\beta=1,2,3,4,5$ the reweighting factor is 1.0, 1.0, 0.984(7), 0.964(13), and 0.972(17) respectively, so that the sign problem remains very mild, at least for these parameter values.

As a further test we also looked at the full subsets for small lattices in three- and four-dimensional QCD, even though the computational cost is huge even for small lattices. We observe with great interest that for $2^3$, $4\times2^2$, and $2^4$ lattices the full subsets always give positive weights, as was verified on samples of 200 random configurations.

\section{Summary}

We have applied subsets generated by center multiplications on temporal links to QCD at nonzero chemical potential and have proven the emerging subset weights to be positive for $N_t=2$ lattices in the strong coupling limit. We have also presented numerical evidence that leads us to conjecture that the positivity persists for larger lattices. Moreover, preliminary results show that the reweighting induced when reintroducing the gauge action is not severe. In a subset measurement of the quark number density the typical Silver Blaze phenomenon is clearly visible.

The number of determinants constituting the full subsets grows exponentially with the number of lattice sites, i.e.\ with inverse temperature and volume, which is a reincarnation of the sign problem. This is why our numerical studies have been restricted to small lattices so far. Two ways out of this situation are conceivable. First, smaller than full subsets improve the reweighting factor considerably -- for which we gave an analytic argument as well as numerical evidence -- and  these subsets are thus helpful for extending the applicability range of reweighting methods.

The second possibility relies on our finding that the full subset weight can be decomposed in a sum of positive terms. Some of their weights simplify considerably, as will also be discussed in a further publication \cite{Bloch:2015zzz}. Moreover, the Grassmann nature of the fermions constrains the combinations of such building blocks. Such constrained systems can typically be simulated using worm algorithms, and will be the subject of future work. 

Open issues are the conjecture about the subset positivity for $N_t\geq 4$ lattices, which still needs to be demonstrated, and the need to investigate how large the sign problem becomes on larger lattices when introducing the gauge action and leaving the strong coupling regime.

\section{Acknowledgments}

FB thanks Björn Wellegehausen and Andreas Wipf for useful discussions. This work has been supported by the DFG (SFB/TRR-55 and BR 2872/6-1).

\appendix

\section{Positivity of the massive subsets}
\label{app:massive}

In this section we extend the positivity proof given in Sec.\ \ref{sec:ferm_proof} to the massive case.

The mass insertions, up to three per site, 
\begin{align}
 &\exp\left(2m\,\bar\psi(x)\psi(x)\right)
 =1+2m\,\bar\psi(x)\psi(x)\nonumber\\
 &+\frac{(2m)^2}{2!}\,(\bar\psi(x)\psi(x))^2+\frac{(2m)^3}{3!}\,(\bar\psi(x)\psi(x))^3\,,
 \label{expm}
\end{align}
break the chiral symmetry, since they connect sites with themselves, and thus modify the positivity argument of Sec.\ \ref{sec:ferm_proof}, which was based on the staggered chirality of the Dirac operator.

For a given subset term let $n$ denote the total number of mass insertions on even sites.  The total number of mass insertions at odd sites is $n$ too\footnote{For $n$ mass insertions on even sites the remaining $(3V/2-n)$ fermions $\psi(x^e)$ must be provided by temporal and spatial hoppings to saturate the Grassmann integrals. These hoppings also contain the same number of $\bar\psi(x^o)$. The same is true for $\bar\psi(x^e)$ and $\psi(x^o)$. In order to saturate the Grassmann integrals on the odd sites, $n$ mass insertions are required there as well.} and thus the partition function only contains even powers of $m$. For the argument it is irrelevant how many insertions come from any given site, the only thing that matters is that the coefficients of all even mass terms in \eqref{expm} have the same sign and those from all odd mass terms too. We denote the mass insertion locations as $x^o_{i}$ and $x^e_{j}$ with $i,j=1,\ldots,n$ and regroup the fermions 
as follows,
\begin{align}
 &\Big[\prod_{i=1}^n \bar\psi(x^o_{i})\psi(x^o_{i})\Big]
  \Big[\prod_{j=1}^n \bar\psi(x^e_{j})\psi(x^e_{j})\Big]\nonumber\\
 &=\Big[\prod_{i=1}^n \bar\psi(x^o_{i})\Big]
  \Big[\prod_{i=1}^n \psi(x^o_{i})\Big]
  \Big[\prod_{j=1}^n \bar\psi(x^e_{j})\Big]
  \Big[\prod_{j=1}^n \psi(x^e_{j})\Big]\nonumber\\
 &=
   \Big[\prod_{i=1}^n \bar\psi(x^o_{i})\Big]
   \Big[\prod_{j=1}^n \psi(x^e_{j})\Big]\cdot
   \Big[\prod_{i=1}^n \psi(x^o_{i})\Big]
   \Big[\prod_{j=1}^n \bar\psi(x^e_{j})\Big]\,,
\end{align}
which comes with a factor $m^{2n}$.
These mass insertions multiply expressions similar to \eqref{sigma_4.17} for the hopping contributions without changing their structure, i.e.\ they remain products of complex conjugate polynomials after Grassmann integration.  Because the even powers of $m$ have positive coefficients and no additional minus signs are picked up by reordering the Grassmann variables (a reordering of $\psi$ is always accompanied by the same reordering of $\bar\psi$) the positivity proof holds as before.

\section{Mixed-meson contribution}
\label{app:Meson-contrib}

In order to simplify the mixed-meson hopping $(u\ub)^2(v\vb)$ we substitute $\psi\equiv \psi(1,\vec x)$, $\varphi\equiv\psi(2,\vec x)$, $U\equiv U_0(1,\vec x)$ and $V=U_0(2,\vec x)$ in \eqref{ubdef}, for simplicity, and expand
\begin{align}
 u^2\vb 
 &= (\bar{\psi}U\psiy)^2(-\bar\psi V^\dagger\varphi) 
 = \bar{\psi}_a \bar{\psi}_b \bar\psi_c U_{aa'} U_{bb'} \psiy_{a'}\psiy_{b'}\psiy_{c'}V^\dagger_{cc'}\nonumber\\
 &= \bar\psi^3 \psiy^3 \epsilon_{abc} \epsilon_{a'b'c'} U_{aa'} U_{bb'} V^\dagger_{cc'}
\label{eq:two_meson_expanded}
 \\
 \ub^2 v &= (-\bar\psiy U^\dagger\psi)^2(\bar\varphi V \psi) 
 = -\bar{\psiy}_a \bar{\psiy}_b \bar\psiy_c U^\dagger_{aa'} U^\dagger_{bb'}\psi_{a'} \psi_{b'} \psi_{c'}V_{cc'} \nonumber\\
 &= - \bar\psiy^3 \psi^3 \epsilon_{abc} \epsilon_{a'b'c'} U^\dagger_{aa'} U^\dagger_{bb'}  V_{cc'} ,
 \label{eq:two_meson_expanded2}
\end{align}
where we also reordered the Grassmannians.
We will also use the following relation for $U \in \text{SU}(3)$,
\begin{align}
 \tfrac12\epsilon_{abc}\epsilon_{a'b'c'} U_{aa'}U_{bb'} = U^\dagger_{c'c} .
 \label{eq:U2Udagger}
\end{align}
Indeed, $U^\dagger=U^{-1}$ in SU(3) and the left hand side of \eqref{eq:U2Udagger} is the matrix of cofactors, which equals $U^{-1}$ transposed, up to the determinant of $U$, which is unity in SU(3).

Substituting \eqref{eq:U2Udagger} in \eqref{eq:two_meson_expanded} and \eqref{eq:two_meson_expanded2} yields
\begin{align}
\begin{aligned}
u^2\vb &= 2 \bar\psi^3 \psiy^3 U^\dagger_{c'c} V^\dagger_{cc'}
= 2 \tr P^\dagger \, \bar\psi^3\varphi^3 ,
\\
\ub^2 v &= - 2 \bar\psiy^3 \psi^3 U_{c'c}  V_{cc'}
= -2\tr P \, \bar\varphi^3\psi^3 ,
\end{aligned}
\end{align}
with Polyakov line $P=UV$. Gathering these results yields
\begin{align}
(u\ub)^2(v\vb) = 4 \tr P \tr P^\dagger \, \bar\psi^3\psi^3 \bar\varphi^3\varphi^3 ,
\end{align}
where the sign vanishes after commuting Grassmann variables.

\section{Temporal hoppings for $2\times2$ lattices}
\label{app:temp_hoppings}

Herein we compute the subsets of temporal hoppings needed in \eqref{sigmaI}. Their definitions \eqref{eq:T_one} as sums of two temporal links can be compactly written as
\begin{align}
 T_x^{\eo,\oe}
 &=(-)^{x+1}\Big[\ep U_0(t^{\eo,\oe},x)+\emu U_0^\dagger(t^{\oe,\eo},x)\Big]\,,\\
&\text{with}\quad t^{\eo}=x,\quad t^{\oe}=(x+1)\,\text{mod}\,2\,.\nonumber
\end{align}
The determinants are computed in the following way 
\begin{align}
 &\det T^{\eo,\oe}_{x} = (-)^{x+1} 
\det\big(\ep \underbrace{U_0(t^{\eo,\oe},x)U_0(t^{\oe,\eo},x)}_{\textstyle P(t^{\eo,\oe},x)}+\,\emu\big) 
\nonumber\\
&\hspace{26mm}\times \underbrace{\det U_0^\dagger(t^{\oe,\eo},x)}_{\textstyle 1}
 \nonumber\\
&= (-)^{x+1} \big(\emum + \emu \tr P(x)+ \ep \tr P(x)^\dagger + \epp\big) \,,
\label{eq:detT_alone}
\end{align}
where the Polyakov loops $P$ only depend on the spatial argument after tracing, and the second equality can be derived for any $P\in\SU(3)$ using its eigenvalues. The product of these determinants at the same spatial position~$x$,
\begin{align}
& \det T^{\eo}_x\det T^{\oe}_x = 
  e^{-6\mu} +e^{-4\mu}\,2\tr P(x) \nonumber\\
 & +e^{-2\mu}\left[2\tr P(x)^\dagger+(\tr P(x))^2\right] 
 + 2(1+|\tr P(x)|^2) \nonumber\\
 &+e^{2\mu}\left[2\tr P(x)+(\tr P(x)^\dagger)^2\right] 
 + e^{4\mu}\, 2\tr P(x)^\dagger + e^{6\mu} \,,
 \label{eq:oned_all}
\end{align}
is nothing but the one-flavor determinant of massless one-dimensional QCD for $\Nt=2$ \cite[Eq.\ (A.6) with $A=2$]{Bloch:2013ara}.

What is needed in \eqref{sigmaI} are total subsets on various products of such determinants. Since the latter depend on the links only through Polyakov loops, it is sufficient to center rotate the $P(x)$, which removes their nonzero triality terms. From \eqref{eq:detT_alone} we obtain the subset as the zero triality projection
\begin{align}
 \sigma\left(\det T^{\eo,\oe}_x\right) &=(-)^{x+1}\,2\cosh(3\mu)\,. 
\end{align}
Note that $T_x^{\eo}$ and $T_x^{\oe}$ depend on the same $P(x)$ such that the subset on the product \eqref{eq:oned_all},
\begin{align}
 \sigma\left(\det T^{\eo}_x\det T^{\oe}_x\right)
 &= e^{-6\mu}+2+2|\tr P(x)|^2+e^{6\mu} \nonumber\\
 &= 4\cosh^2(3\mu)+2|\tr P(x)|^2\,,
\end{align}
is not the product of individual subsets. Again, this expression agrees with the subset weight for the massless case in 0+1 dimension, cf.~Eq.~\eqref{eq:oned_result_massless}. Finally, the subsets on different $x$ factorize, giving
\begin{align}
&\sigma\left(\det T^{\eo}_1\det T^{\eo}_2\right) 
 = \sigma\left(\det T^{\oe}_1\det T^{\oe}_2\right) =-4\cosh^2(3\mu) ,
\nonumber\\
 &\sigma\left(\det T^{\eo}_1\det T^{\eo}_2\det T^{\oe}_1\det T^{\oe}_2\right)  \label{eq:det_subsetted}\\
  &=\left(4\cosh^2(3\mu)+2|\tr P(1)|^2\right)\times\left(P(1) \to P(2)\right) \,.
\nonumber
\end{align}

\bibliography{z3pos}

\end{document}